\documentclass[conference]{IEEEtran}
\IEEEoverridecommandlockouts
\usepackage{cite}
\usepackage{amsmath,amssymb,amsfonts}
\usepackage{algorithmic}
\usepackage{algorithm}
\usepackage{graphicx}
\usepackage{textcomp}
\usepackage{xcolor}
\def\BibTeX{{\rm B\kern-.05em{\sc i\kern-.025em b}\kern-.08em
		T\kern-.1667em\lower.7ex\hbox{E}\kern-.125emX}}

\usepackage{booktabs}

\usepackage{tabularx}
\usepackage{array}
\usepackage{graphicx}
\usepackage{float}
\usepackage{subfig}

\makeatletter
\newcommand{\linebreakand}{
\end{@IEEEauthorhalign}
\hfill\mbox{}\par
\mbox{}\hfill\begin{@IEEEauthorhalign}
}
\makeatother

\begin{document}
\title{Extreme Scenario Characterization for High Renewable Energy Penetrated Power Systems over Long Time Scales\\
}
\author{\IEEEauthorblockN{Kai Kang, Feng Liu*}
\IEEEauthorblockA{\textit{Department of Electrical Engineering,}\\
	\textit{ Tsinghua University} \\
	Beijing, China \\
kk21@tsinghua.org.cn; lfeng@tsinghua.edu.cn}
\and
\IEEEauthorblockN{Yifan Su}
\IEEEauthorblockA{\textit{School of Electrical Engineering,}\\
	\textit{Chongqing University} \\
	Chongqing, China \\
	suyf@cqu.edu.cn
	}
\and
\IEEEauthorblockN{Zhaojian Wang}
\IEEEauthorblockA{\textit{Department of Automation,}\\
	\textit{Shanghai Jiao Tong University} \\
	Shanghai, China \\
wangzhaojian@sjtu.edu.cn}
}
\maketitle
\begin{abstract}
Power systems with high renewable energy penetration are highly influenced by weather conditions, often facing significant challenges such as persistent power shortages and severe power fluctuations over long time scales. This paper addresses the critical need for effective characterization of extreme scenarios under these situations. First, novel risk indices are proposed to quantify the severity of continuous power shortages and substantial power fluctuations over long-term operations. These indices are independent of specific scheduling strategies and incorporate the system's resource regulation capabilities. By employing a filtering-based approach, the proposed indices focus on retaining key characteristics of continuous power shortages and fluctuation events, enabling the identification of extreme scenarios on long time scales. Secondly, an extreme scenario generation method is developed using Gaussian mixture models and sequential Monte Carlo simulation. Especially, this method periodically evaluates the severity of generated scenarios based on the defined risk indices, retaining extreme scenarios while discarding less critical ones. Finally, case studies based on real-world data demonstrate the efficacy of the proposed method. The results confirm that integrating the identified extreme scenarios significantly enhances the system's ability to ensure long-term security and reliability under high renewable energy penetration.
\end{abstract}


\section{Introduction}
The increasing integration of renewable energy sources into power systems has established a closer coupling between power systems and meteorological conditions \cite{chen2019probabilistic}. This growing dependency introduces significant uncertainties on both the supply (generation) and demand (load) sides, posing substantial challenges to maintaining the power balance, particularly over long-term time scales (e.g., quarterly, annual). On these horizons, renewable energy variability and persistent supply-demand imbalances have become more pronounced. For instance, in February 2021, record-breaking low temperatures across central regions of the United States led to widespread shutdowns of renewable energy systems and a sudden surge in demand, resulting in prolonged power outages across Texas \cite{busby2021cascading}. Similarly, in August 2022, Sichuan, China, faced extreme heatwaves and droughts, necessitating power curtailment measures for 11 consecutive days \cite{zhou2023novel}. Such events underscore the urgency of developing effective methods to characterize extreme scenarios for long-term power system assessment, planning, operation, and maintenance under high renewable energy penetration \cite{bennett2021extending}.

Various approaches for extreme scenario characterization have been proposed, including methods based on historical data screening, probability density tail risk assessment, and worst-case scenario analysis within uncertainty intervals \cite{li2014risk,abdin2019modeling,kang2023day,chow2015risk}. However, these methods exhibit notable limitations. Historical data screening is often inadequate for identifying future extreme events, as many such scenarios are unprecedented. Tail risk-based probability density methods evaluate risks at the tail ends of probability distributions. However, the long-term scales encompass a large number of operational periods with complex, coupled uncertainties, making it difficult to accurately measure and assess tail risks. Meanwhile, worst-case scenario approaches define extreme scenarios within predefined uncertainty bounds. On long time scales, however, determining precise uncertainty intervals becomes highly challenging. Furthermore, extreme scenarios may occur outside these preset intervals, rendering the method insufficient for capturing truly representative worst-case conditions.

In light of these challenges, index-based generation and screening of extreme scenarios have gained attention. Widely used risk assessment metrics, such as loss-of-load probability (LOLP) and expected energy not supplied (EENS) are instrumental in evaluating power system performance under extreme conditions \cite{hu2021decision}.
Typically, these indicators are applied to scenarios generated through methods such as Monte Carlo simulations \cite{chen2007nonparametric} or Copula functions \cite{salvadori2016multivariate}. By incorporating the grid’s operational strategies, these indicators facilitate the identification of extreme scenarios. However, this approach is inherently dependent on the specific operational strategy in use. Consequently, any change in the strategy can lead to a different set of screened extreme scenarios, reducing the robustness of the results. 
Literature \cite{perera2020quantifying}  introduces risk assessment indicators based on temperature to identify scenarios of extreme cold year and extreme warm year, providing insights into the annual operational conditions of renewable energy generation and load demand. However, this approach does not account for the impact of power fluctuations on both the supply and demand sides. 
Literature \cite{liang2022climate} considers the influence of renewable energy power fluctuations but neglects the inclusion of the power grid’s regulatory resource endowments in the risk indicators. Moreover, on long-term scales such as quarterly and annual, the unique challenges, including persistent power shortages and power fluctuations require greater attention. Existing indicators fail to adequately capture these persistent shortages and fluctuations, highlighting the need for refined risk assessment methods that can better address long-term operational challenges under high renewable energy penetration.

To address these gaps, this paper proposes a novel methodology for characterizing extreme scenarios in power systems with high renewable energy penetration over long time scales. Firstly, considering the persistent power shortages and fluctuations on both the supply and demand sides as well as the regulating resource capabilities of the power system, we propose new risk indices to characterize it. These indices are designed to be independent of specific dispatch strategies, enabling a focused analysis of continuous power supply shortages and fluctuation events. Secondly, building on the proposed indices, we develop an extreme scenario generation method using the sequential Monte Carlo. This method enables that scenarios with the greatest potential impact on the power system over long time scales to be identified and retained during the generation process. Finally, using real-world data, the proposed indices and scenario screening method are validated to demonstrate their effectiveness in ensuring the safe and reliable operation of power systems under high renewable energy penetration.

\section{Risk index for extreme scenarios over long time scales}
\subsection{Outline}
In this section, we introduce novel risk indices to characterize the risk of extreme scenarios in high renewable energy penetrated power systems.

The power system consists of the synchronous generator (SG), renewable generation (RG), long-term energy storage (LTES), short-term energy storage (STES), load demand (LD), and transmission line (TL). 
Each type of device is represented by a set $\Omega_\bullet$, and the number of devices in the set is $\left|\Omega_\bullet\right|$, where $\bullet$ can be expressed as SG, RG, LTES, STES, LD, and TL, respectively. The mathematical model of the above component can be referenced from \cite{kang2024enforcing} and \cite{jiang2022renewable}. The set of periods over long time scale is denoted as $\Omega_\mathrm{T}$, and the number of periods is $N_{\mathrm{T}}$. 

Let $\Omega_\mathrm{EXT}$ represent the set of extreme scenarios, and $N_\mathrm{EXT}$ their number. The $e$-th ($e \in {\Omega _{{\rm{EXT}}}}$) extreme scenario, $\overline P_e^{{\rm{EXT}}}$, is defined as:
\begin{equation}
	\notag
	\overline P_e^{{\rm{EXT}}} = \left\{\left\{\overline P_{r,t,e}^{\rm{RG}}\right\} _{r \in \Omega_\mathrm{RG} },  \, \, \left\{ \overline P_{d,t,e}^{\rm{LD}} \right\} _{d \in \Omega_\mathrm{LD}}   \right\}_{t = 1}^{{N_{\mathrm{T}}}},
\end{equation}
where $\overline P_{r,t,e}^{\mathrm{RG}}$ and $\overline P_{d,t,e}^{\mathrm{LD}}$ represent the available renewable generation and load demand, respectively, for extreme scenario $e$ in period $t$.

\subsection{Risk characterization for power shortage}
In high renewable energy penetrated power systems, uncertainties in renewable generation and load occur across multiple time scales, from minute-level fluctuations to seasonal variations. These uncertainties pose significant challenges to grid security, particularly under scenarios of prolonged low renewable output and sustained high load demand. For an extreme scenario $\overline P_e^{{\rm{EXT}}}$ at period $t \in \Omega_\mathrm T$, we define the power shortage index $\mathrm{IPS}_e(t)$ as:
\begin{equation}
	\setlength\abovedisplayskip{0pt}
	\setlength\belowdisplayskip{0pt}
	\label{eq:ID_power_shortage}
	\mathrm{IPS}_e(t) = \sum\limits_{\tau = t}^{t+\sigma}\mathrm{sgn}[\mathrm{PS}_e(\tau)]\cdot\left[\mathrm{PS}_e(t)\right]^2
\end{equation}
where $\mathrm{sgn}[x]$ represents the Signum function, which equals 1 if $x > 0$, and 0 otherwise. Here, $\mathrm{PS}_e(t)$ denotes the power shortages under extreme scenario $e$, given by:
\begin{equation}
	\setlength\abovedisplayskip{0pt}
	\setlength\belowdisplayskip{0pt}
	\label{eq:ID_power_shortage_2}
	\begin{aligned}
	&\mathrm{PS}_e(t) = \\
	&\max\left(\sum\limits_{d \in \Omega_\mathrm{LD}} \overline P_{d,t,e}^{\rm{LD}} - \alpha^\mathrm{SG}\sum\limits_{g \in {\Omega _\mathrm{SG}}} \overline{P}_{g}^\mathrm{SG}  - \sum\limits_{r \in {\Omega _\mathrm{RG}}} \overline P_{r,t,e}^\mathrm{RG},0 \right)
	\end{aligned}
\end{equation}
where $\overline{P}_{g}^\mathrm{SG}$ is the maximum generation capacity of synchronous generator $g \in \Omega _\mathrm{SG}$, and $\alpha^\mathrm{SG} \in [0,1]$ is the preset coefficient. When the combined output of synchronous generators and renewable energy sources satisfies load demand, $\mathrm{PS}_e(t) = 0$. 

The parameter $\sigma$ defines the bandwidth of characterizing continuous power loss, accounting for future periods $[t+1, t+\sigma]$ where power shortages occur: If sustained mismatches are detected, the power shortage index $\mathrm{IPS}_e(t)$ increases accordingly. 

The proposed power shortage index directly evaluates extreme scenarios without relying on specific scheduling strategies. It incorporates the resource endowment of synchronous generators and captures the effects of renewable generation and load demand. By employing the Signum function and the bandwidth parameter $\sigma$, this index effectively characterizes the risk of prolonged power shortages.

\subsection{Risk characterization for persistent power fluctuations}
Over long time scales spanning multiple operational periods, the combined power fluctuations of renewable energy and load demand in adjacent periods may exceed the system's ramping capability, leading to potential power loss. To quantify this risk, we define the power fluctuation index $\mathrm{IPF}_e(t)$ for extreme scenario $e$ in period $t \in \Omega_\mathrm T$: 
\begin{equation}
	\setlength\abovedisplayskip{0pt}
	\setlength\belowdisplayskip{0pt}
	\label{eq:ID_power_fluctuation}
	\mathrm{IPF}_e(t) = \sum\limits_{\tau = t}^{t+\sigma}\mathrm{sgn}[\mathrm{PF}_e(\tau)]\cdot\left[\mathrm{PF}_e(t)\right]^2
\end{equation}
where $\mathrm{PF}_e(t)$ represents the magnitude of power fluctuation for scenario $e$, calculated as: 
\begin{equation}
	\setlength\abovedisplayskip{0pt}
	\setlength\belowdisplayskip{0pt}
	\label{eq:ID_power_fluctuation_2}
	\begin{aligned}
		\mathrm{PF}_e(t) = \max\left[\mathrm{RU}_e^\mathrm{PF}(t), \mathrm{RD}_e^\mathrm{PF}(t)\right],
	\end{aligned}
\end{equation}

In this expression, $\mathrm{RU}_e^\mathrm{PF}(t)$ and $\mathrm{RD}_e^\mathrm{PF}(t)$ represent the ramp-up and ramp-down fluctuation amounts, respectively, given by:
\begin{equation}
	\setlength\abovedisplayskip{0pt}
	\setlength\belowdisplayskip{0pt}
	\label{eq:ID_power_fluctuation_3}
	\begin{aligned}
		\mathrm{RU}_e^\mathrm{PF}(t) =	
		\max \left\{ {\begin{array}{*{20}{c}}
				{\sum\limits_{d \in \Omega_\mathrm{LD}}(\overline P_{d,t,e}^{\rm{LD}} - \overline P_{d,t-1,e}^{\rm{LD}}) }\\
				{ +\sum\limits_{r \in {\Omega _\mathrm{RG}}}( \overline P_{r,t-1,e}^\mathrm{RG} -\overline P_{r,t,e}^\mathrm{RG}) }\\
				{-\sum\limits_{g \in {\Omega _\mathrm{SG}}} \mathrm{RU}_{g}^\mathrm{SG} }
			\end{array},0} \right\}
	\end{aligned}
\end{equation}
\begin{equation}
	\setlength\abovedisplayskip{0pt}
	\setlength\belowdisplayskip{0pt}
	\label{eq:ID_power_fluctuation_4}
	\begin{aligned}
		\mathrm{RD}_e^\mathrm{PF}(t) =	
		\max \left\{ {\begin{array}{*{20}{c}}
				{\sum\limits_{d \in \Omega_\mathrm{LD}}
				(\overline P_{d,t-1,e}^{\rm{LD}} - \overline P_{d,t,e}^{\rm{LD}}) }\\
				{ +\sum\limits_{r \in {\Omega _\mathrm{RG}}}( \overline P_{r,t,e}^\mathrm{RG} - \overline P_{r,t-1,e}^\mathrm{RG}) }\\
				{-\sum\limits_{g \in {\Omega _\mathrm{SG}}} \mathrm{RD}_{g}^\mathrm{SG} }
			\end{array},0} \right\}
	\end{aligned}
\end{equation}
where $\mathrm{RU}_{g}^\mathrm{SG}$ and $\mathrm{RD}_{g}^\mathrm{SG}$ represent the maximum ramp-up and ramp-down capacities of synchronous generator $g \in \Omega _\mathrm{SG}$. Similar to the power shortage index $\mathrm{IPS}_e(t)$, the parameter $\sigma$ in $\mathrm{IPF}_e(t)$ highlights the effects of continuous power fluctuations on the power system.

\subsection{Risk index for extreme scenarios over long time scales}
To evaluate extreme scenarios comprehensively, we combine the power shortage index $\mathrm{IPS}_e(t)$ and the power fluctuation index $\mathrm{IPF}_e(t)$ into a unified long-term risk index $\mathrm{ILT}_e$:
\begin{equation} 
	\setlength\abovedisplayskip{0pt}
	\setlength\belowdisplayskip{0pt}
	\label{eq:ILT}
	\mathrm{ILT}_e = \sum\nolimits_{t \in \Omega_\mathrm{T}}\left[\theta\cdot\mathrm{IPS}_e(t) + (1-\theta)\cdot \mathrm{IPF}_e(t)\right]
\end{equation}
where $\theta$ is the weight coefficient that balances the contributions of the two indices. 

The index $\mathrm{ILT}_e$ provides a holistic measure of the severity of extreme scenarios over long time scales. A higher $\mathrm{ILT}_e$ value indicates more significant risks associated with persistent power shortage and power fluctuations, reflecting the scenario’s impact on system stability and security.

\section{Extreme scenario generation method using risk indices and sequential Monte Carlo}
We propose a method for generating extreme scenarios based on historical data. Let the set and number of historical scenarios, including renewable generation and load data, be denoted as $\Omega_\mathrm{HST}$ and $\left|\Omega_\mathrm{HST}\right|$, respectively. The $h$-th historical scenario is defined as:
\begin{equation}
	\notag
	\overline P_h^{{\rm{HST}}} = \left\{ \left\{\overline P_{r,t,h}^{\rm{RG}}\right\} _{r \in \Omega_\mathrm{RG} },  \, \, \left\{ \overline P_{d,t,h}^{\rm{LD}} \right\} _{d \in \Omega_\mathrm{LD}}   \right\}_{t = 1}^{{N_{\rm{T}}}},
\end{equation}
where $\overline P_{r,t,h}^{\rm{RG}}$ and $ \overline P_{d,t,h}^{\rm{LD}}$ represent the renewable generation and load demand, respectively, for historical scenario $h$ in period $t$.

Historical data serves as a foundation for generating extreme scenarios. However, the process poses several challenges:

1. Characterizing uncertainty: The uncertainty in renewable generation and load cannot be accurately described using a single probability distribution.

2. Time-dependent coupling: The uncertain powers across adjacent time periods are interdependent, meaning the value in period $t-1$ influences the distribution in period $t$.

3. Efficiency: Extreme scenarios must be generated efficiently, with less computational overhead.

To address these challenges, we utilize a Gaussian Mixture Model (GMM)-based Sequential Monte Carlo (SMC) method, incorporating importance sampling and the proposed risk indices for generating and screening extreme scenarios. 

The framework is detailed in Algorithm \ref{alg:scenario_generation}. 
GMM approximates complex probability distributions by tuning the number of components and their parameters. This overcomes the limitations of single-distribution models for representing uncertainties. 
The sequential Monte Carlo captures temporal dependencies between adjacent time periods, enabling that the coupling of uncertain power across periods is represented. The importance sampling prioritizes scenarios that are more likely to result in extreme events, and the proposed risk indices enables for evaluating the scenarios' severity, allowing for rapid screening of extreme cases.

 \begin{algorithm}[htb] 
 	\caption{Extreme scenario generation using risk index and sequential Monte Carlo.} 
 	\label{alg:scenario_generation} 
 	\textbf{Step 1:} 
 	Obtain the historical scenario set $\Omega_\mathrm{HST}$ and power system parameters. Set the risk index screening interval $\pi_s$. Set the sample number of extreme scenarios $N_\mathrm{EXT}^\mathrm{SAM}$, where $N_\mathrm{EXT}^\mathrm{SAM} = \left(\frac{N_{\mathrm{T}}}{\pi_s} + 1 \right) \cdot N_\mathrm{EXT}$. \\
 	\textbf{Step 2:} 
 	Set the initial period index $t \leftarrow 1$. Construct the Gaussian mixture distribution for $\overline P_{r,t,h}^{\rm{RG}}$ and $ \overline P_{d,t,h}^{\rm{LD}}$ ($h \in \Omega_\mathrm{HST}$, $r \in \Omega_\mathrm{RG}$, $d \in \Omega_\mathrm{LD}$). Sample from these distributions to obtain $\overline P_{r,t,e}^{\rm{RG}}$ and $ \overline P_{d,t,e}^{\rm{LD}}$  ($e \in \left[1,N_\mathrm{EXT}^\mathrm{SAM}\right]$, $r \in \Omega_\mathrm{RG}$, $d \in \Omega_\mathrm{LD}$). \\
 	\textbf{Step 3:} 
 	Update $t \leftarrow t+1$. 
 	Generate a two-dimensional Gaussian mixture distribution based on $\overline P_{r,t-1,h}^{\rm{RG}}$ and $\overline P_{r,t,h}^{\rm{RG}}$ ($h \in \Omega_\mathrm{HST}$, $r \in \Omega_\mathrm{RG}$, $d \in \Omega_\mathrm{LD}$). Using the conditional probability distribution at the sampled point $\overline P_{r,t-1,e}^\mathrm{RG}$, and then obtain $\overline P_{r,t,e}^\mathrm{RG}$ ($e \in \left[1,N_\mathrm{EXT}^\mathrm{SAM}\right]$, $r \in \Omega_\mathrm{RG}$) using importance sampling. Apply a similar process to generate $ \overline P_{d,t,e}^{\rm{LD}}$ ($e \in \left[1,N_\mathrm{EXT}^\mathrm{SAM}\right]$, $d \in \Omega_\mathrm{LD}$). \\
 	\textbf{Step 4:} If $t\ge N_{\mathrm{T}}$, proceed to Step 5.\\
 	If $t = n\cdot \pi_s$ ($n \in \mathbb{N}^+$), calculate the risk index $\mathrm{ILT}_e$ for all extreme scenarios.
 	Delete the $N_\mathrm{EXT}$ scenarios with the lowest $\mathrm{ILT}_e$ values, update the total sample count as $N_\mathrm{EXT}^\mathrm{SAM} \leftarrow N_\mathrm{EXT}^\mathrm{SAM} - N_\mathrm{EXT}$, and return to Step 3. \\
 	Otherwise, return to Step 3.\\
 	\textbf{Step 5:} End the iterations. 
 	Calculate the risk index $\mathrm{ILT}_e$ for 
 	all remaining extreme scenarios. 
 	select the top $N_\mathrm{EXT}$ scenarios with the highest $\mathrm{ILT}_e$ values to
 	form the set $\Omega_\mathrm{EXT}$. Output the final set $\Omega_\mathrm{EXT}$.
 \end{algorithm}

\section{Case Studies}
\subsection{Set-Up}
Germany has experienced rapid development in renewable energy. By 2023, the total installed renewable energy capacity in Germany has reached approximately 170 GW, with a relatively balanced contribution from wind and photovoltaic power. Concurrently, the country's peak load stands at around 76 GW.
\begin{figure*}[htb]
	\centering
	\subfloat[Wind generation]{\label{Fig:Scenario:a}
		\includegraphics[width=5.7cm]{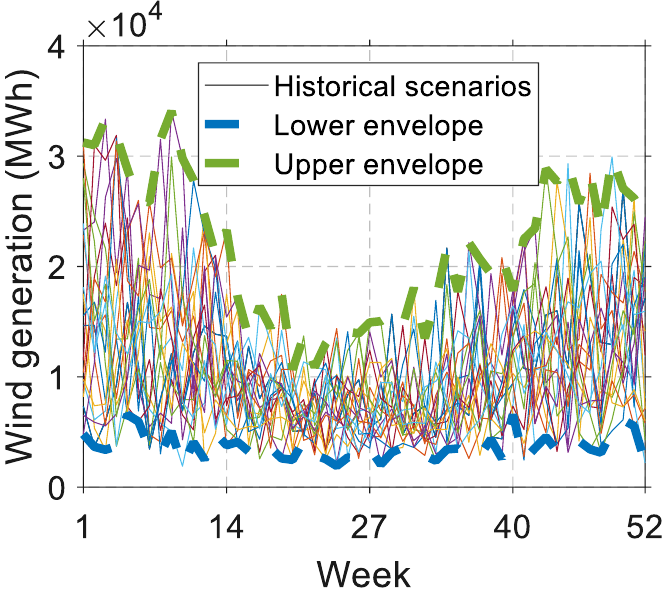}} 
	\hfill
	\subfloat[PV generation]{\label{Fig:Scenario:b}
		\includegraphics[width=5.7cm]{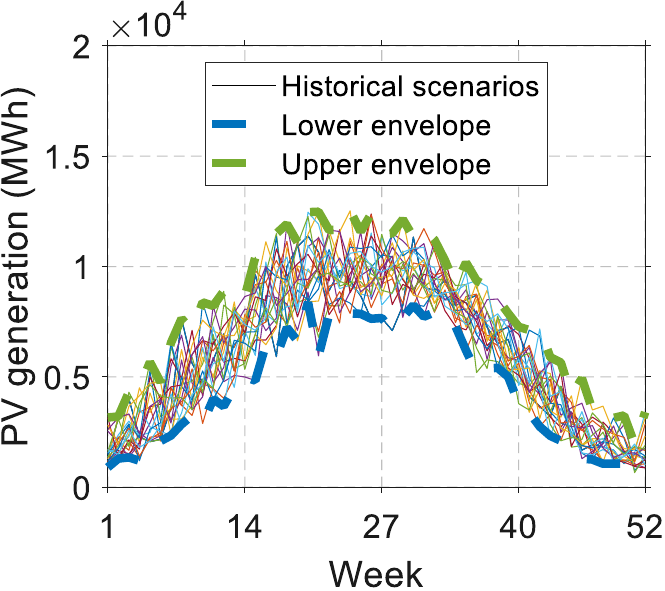}}
	\hfill
	\subfloat[Load demand]{\label{Fig:Scenario:c}
		\includegraphics[width=5.7cm]{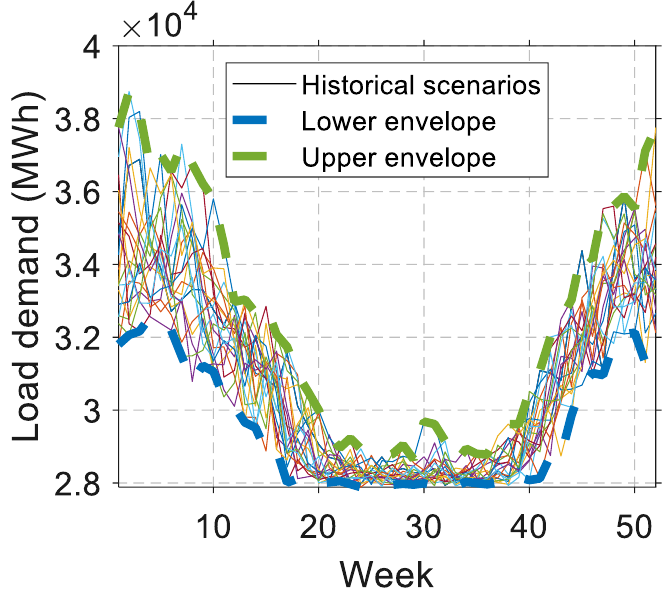}}	
	\caption{Annual renewable energy generation curve and load demand.}
	\label{Fig:Scenarios} 
\end{figure*}
\begin{figure}[htb]
	\centering
	\subfloat[Load mismatch $\mathrm{PS}_e(t)$ and power fluction $\mathrm{PF}_e(t)$]{\label{Fig:Index_2012a}
		\includegraphics[width=8cm]{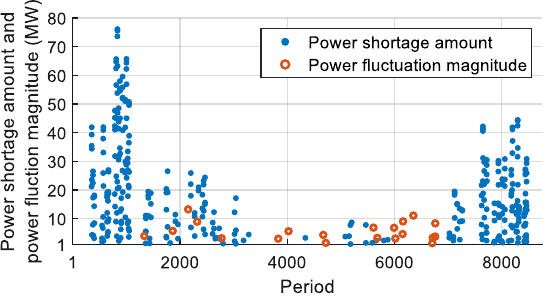}}\\ 
	\subfloat[Values of power shortage and fluctuation indices]{\label{Fig:Index_2012b}
		\includegraphics[width=8cm]{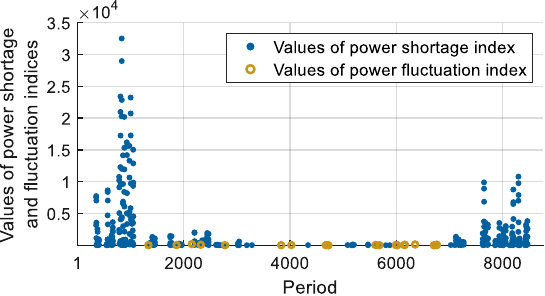}}\\
	\caption{Index values for the historical scenario in 2012.}
	\label{Fig:Index_2012} 
\end{figure} 

Using this context, we construct a high-renewable penetration power system case based on Germany's renewable energy generation and load data spanning 1980 to 2019 \cite{staffell2023global}. This case facilitates the characterization and analysis of extreme scenarios, including wind and photovoltaic installed capacities of 300 MW each, a 200 MW synchronous generator with an hourly ramp rate of 60 MW, and an average peak load of 264 MW over the past 40 years. Fig. \ref{Fig:Scenarios} illustrates the annual renewable energy generation curve and load demand for this system. The corresponding dataset is publicly available at \cite{renewables_ninja}.

The above data is utilized in the subsequent sections to validate the effectiveness of the proposed extreme scenario characterization method. The following parameter settings are applied, the coefficient $\alpha^\mathrm{SG}$ in equation \eqref{eq:ID_power_shortage_2} is set to 0.95, the weight coefficient $\theta$ in equation \eqref{eq:ILT} is set to 0.5, and the number of extreme scenarios, $ N_\mathrm{EXT}$, is set to 100.

\subsection{Risk characterization for continuous power shortage and power fluctuation}
To demonstrate the effectiveness of the proposed risk indicators, we use the renewable energy and load data from 2012 as an example.

Fig. \ref{Fig:Index_2012}\subref{Fig:Index_2012a} reveals continuous power shortages around periods 1000 and 8000, while significant power fluctuations are observed around periods 2000, 4000, and 6000. Based on the proposed power shortage index $\mathrm{IPS}_e(t)$ and power fluctuation index $\mathrm{IPF}_e(t)$, the continuous shortages and changes of power in the corresponding periods are highlighted in the indices, as shown in Fig. \ref{Fig:Index_2012}\subref{Fig:Index_2012b}. These results underscore the superior capability of the proposed indices in characterizing continuous power shortage and power fluctuation events.

\subsection{Improving system robustness using the proposed scenario generation method}
Planning long-term energy storage is crucial for ensuring the secure operation of power systems. Using the proposed Algorithm \ref{alg:scenario_generation}, extreme scenarios are generated and ranked by their risk levels. Then, a subset of the most severe scenarios is selected and incorporated into the scenario set for energy storage planning.

\begin{figure}[htb]
	\centering
	\subfloat[Power capacity]{\label{Fig:LDES_planninga}
		\includegraphics[width=8cm]{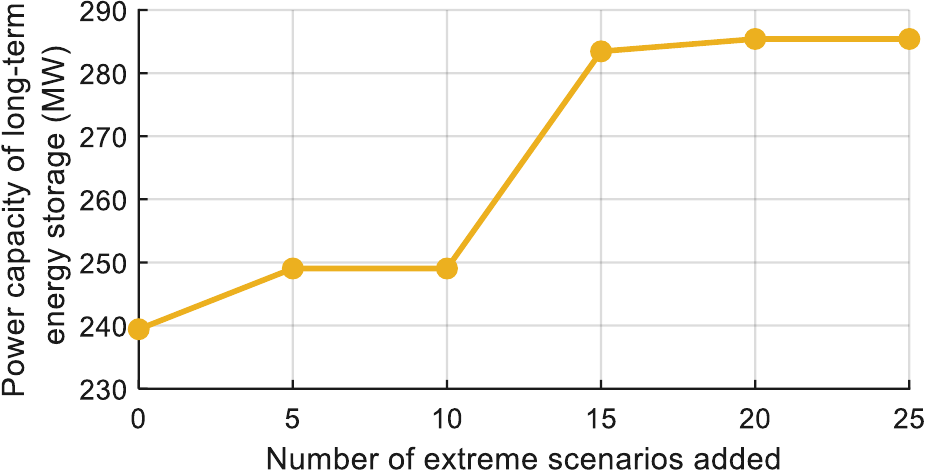}}\\ 
	\subfloat[Energy capacity]{\label{Fig:LDES_planningb}
		\includegraphics[width=8.2cm]{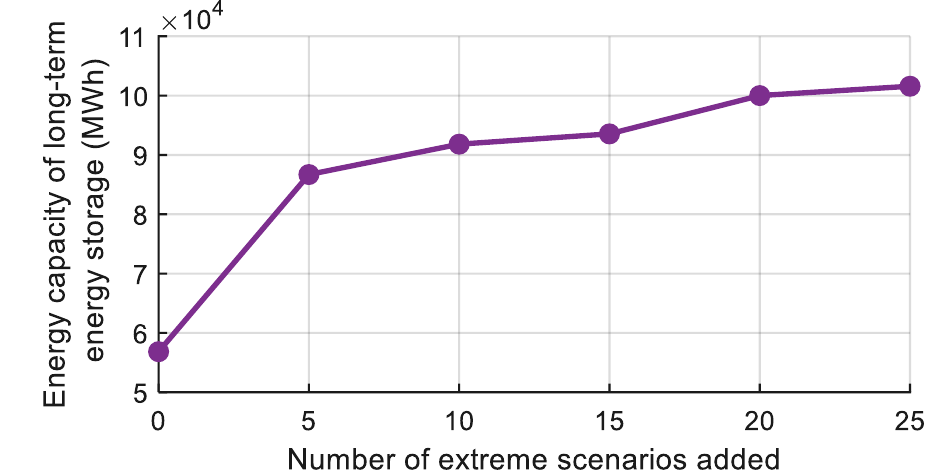}}\\
	\caption{Power capacity and energy capacity planning results of Long-term energy storage.}
	\label{Fig:LDES_planning} 
\end{figure}

Fig. \ref{Fig:LDES_planning} presents the planning results for the power capacity and energy capacity of the long-term energy storage. 

Without incorporating extreme scenarios, the planned power and energy capacities of the long-term energy storage remain relatively low. However, when extreme scenarios are considered, these capacities increase significantly, reflecting the system's enhanced ability to handle adverse conditions. In this case, adding more than 15 extreme scenarios results in a diminished growth rate in the planned long-term energy storage capacities, indicating a saturation effect.

This case highlights the critical role of extreme scenarios in shaping energy storage planning and ensuring the long-term operational safety of the system. Furthermore, it demonstrates that the proposed risk indicators and scenario generation method effectively enable decision-makers to optimize flexible resource allocation with a relatively small number of extreme scenarios.

\section{Conclusions}
This work addresses the critical challenge of managing high renewable energy penetration in power systems, focusing on the persistent issues of power supply shortages and severe fluctuations over long time scales. 
First, innovative risk indices are introduced to quantify the severity of continuous power shortages and substantial fluctuations. These indices are independent of specific scheduling strategies and incorporate the system’s resource regulation capabilities, effectively capturing long-term risks. A filtering-based approach ensures that key characteristics of such extreme events are preserved, enabling accurate identification of high-risk scenarios. Secondly, a scenario generation method based on Gaussian mixture models and sequential Monte Carlo simulation is proposed. This method periodically evaluates scenario severity based on the proposed risk indices, retaining extreme scenarios and eliminating less critical ones, ensuring computational efficiency while maintaining focus on scenarios most relevant to long-term system security.
Finally, comprehensive case studies using real-world data validate the proposed methods. The results highlight the effectiveness of the generated extreme scenarios in bolstering system robustness against continuous power shortages and fluctuations. Integrating these scenarios significantly enhances the system’s ability to maintain long-term reliability and security under high renewable energy penetration.



\ifCLASSOPTIONcaptionsoff
\newpage
\fi
\bibliographystyle{IEEEtran}
\bibliography{IEEEabrv,mybib}

\end{document}